# Few-Layer MoS$_2$/a-Si:H Heterojunction pin-Photodiodes for extended Infrared Detection

Andreas Bablich [1], Daniel S. Schneider [2], Paul Kienitz [1], Satender Kataria [3], Stefan Wagner [2], Chanyoung Yim [1,4], Niall McEvoy [5], Olof Engstrom [2], Julian Müller [1], Yilmaz Sakalli [1], Benjamin Butz [1], Georg S. Duesberg [4,5], Peter Haring Bolívar [1] and Max C. Lemme [2,3]

[1] University of Siegen, School of Science and Technology, Hölderlinstr. 3, Siegen, 57076, Germany

[2] Advanced Microelectronic Center Aachen (AMICA), AMO GmbH, Otto-Blumenthal-Str. 25, 52074 Aachen, Germany

[3] Chair for Electronic Devices, Faculty of Electrical Engineering and Information Technology, RWTH Aachen University, Otto-Blumenthal-Str. 25, 52074 Aachen, Germany

[4] Institute of Physics, EIT 2, Faculty of Electrical Engineering and Information Technology, Universität der Bundeswehr München, Werner-Heisenberg-Weg 39, 85577 Neubiberg, Germany

[5] School of Chemistry and AMBER, Trinity College Dublin, Dublin 2, Ireland

E-Mail: andreas.bablich@uni-siegen.de

Phone: +49 (0) 271 – 740 4748

Fax: +49 (0) 271 – 740 2410

**Abstract**

Few-layer molybdenum disulfide (FL-MoS$_2$) films have been integrated into amorphous silicon (a-Si:H) pin-photodetectors. To achieve this, vertical a-Si:H photodiodes were grown by plasma-enhanced chemical vapor deposition (PE-CVD) on top of large-scale synthesized and transferred homogeneous FL-MoS$_2$. This novel detector array exhibits long-term stability (more than six month) and outperforms conventional silicon-based pin-photodetectors in the infrared range (IR, $\lambda = 2120$ nm) in terms of sensitivies by up to ~50 mAW$^{-1}$. Photodetectivities of up to ~$2 \cdot 10^{10}$ Jones and external quantum efficiencies of ~3 % are achieved. The detectors further feature the additional functionality of bias-dependent responsivity switching between the different spectral ranges. The realization of such scalable detector arrays is an essential step towards pixelated and wavelength-selective sensors operating in the IR spectral range.

Photodetectors play significant roles in fields of optical communication [1], imaging [2], night vision [3], safety and security [4], environmental-/agricultural sensing [5], industrial process monitoring [6], in the food industry [7] and, exceptionally, in fields of research [8]. Established photodetectors for the visible (VIS) regime predominantly consist of crystalline or amorphous silicon (c-Si, a-Si:H) as the absorber due to the maturity and availability of silicon technology. Near-infrared (NIR) (1-2 µm) and mid-IR (2-5 µm) photodetectors are typically based on group IV (Ge) or III-IV semiconductors (e.g., InGaAs, InAsSb, etc.) [9], which necessitate demanding and thus more cost-/energy intensive thin-film fabrication techniques, typically Metal-Organic Chemical Vapor Deposition. Mercury cadmium telluride (HgCdTe) is photosensitive at wavelengths above 2 µm when operated at cryogenic temperatures [10]. Disadvantageous, the noise level of narrow bandgap detectors, like HgCdTe (< 0.1 eV [11]), increases drastically with temperature due to the thermal excitation of charge carriers. Aside from the novel photodetector architecture presented in this work, quantum dot (e.g. colloidal [12] or type-II superlattices) or transition metal oxide based photodetectors, (e.g. VO$_2$/V$_2$O$_5$ heterostructures [13]) might allow reasonable (N)IR detection at higher operating temperatures in the future due to 3D quantum confinement effects.

As applications advance, photoresponse in a combined optical range, here shown from VIS to mid-IR wavelengths, with the capabilities for panchromatic or selective detection (voltage-dependent sensitivity) are desired [14]. In this work, we demonstrate significantly extended photoresponse in well-established a-Si:H photodetectors towards the infrared by the integration of a large-scale 2D semiconductor, the transition metal dichalcogenide (TMD) molybdenum disulfide (MoS$_2$). Even though the bandgap (E$_g$) of bulk MoS$_2$ is only slightly smaller than that of the a-Si:H, a distinct responsivity in the NIR range is observed. Such considerable absorption up to 5 µm has been reported for isolated large-scale MoS$_2$ and depends on its thickness and crystallinity [15].

Although optoelectronic devices based on exfoliated mono- or multilayer MoS$_2$ flakes have been demonstrated to show good photodetection capabilities [16,17], they lack scalability due to the small flake size in the µm range and tedious manual handling. Following the efforts of scalable chemical vapor deposited (CVD) graphene films [18–20], the large-area growth of MoS$_2$ and other TMDs has also been demonstrated using CVD and related techniques [21–24] such as thermally assisted conversion (TAC) of molybdenum (Mo) thin films [25]. The transfer of such grown films is also challenging, but recent progress is quite promising [26]. Beyond material growth and transfer, the large-scale integration of TMDs into semiconductor process technology faces another challenge: the controlled and homogeneous nucleation and growth of functional layers on top of the TMD. Here, few-layer molybdenum disulfide (FL-MoS$_2$) was successfully and reproducibly integrated on a centimeter-scale on two different chips with a conventional thin-film deposition process of a-Si:H pin-photodiodes. The resulting detector arrays can be tuned in their photoresponse from the VIS to the IR range depending on the applied bias voltage. The a-Si:H effectively passivates the 2D MoS$_2$ and provides long-term device stability of at least six month (cf. *Figure S5-S7*). The device technology and fabrication described in this work is generally scalable and suitable for standardized deposition techniques, provided that the transfer of 2D materials becomes an automated process technology.

*Figure 1: a) left: photograph of chips with MoS$_2$/a-Si:H and reference photodiodes, right: microphotograph of circular 200 μm MoS$_2$/a-Si:H detectors, b) fabrication scheme including MoS$_2$ integration, c) cross-sectional STEM micrograph with corresponding EDS element maps, d) HRTEM detail of embedded MoS$_2$ thin film with corresponding fast Fourier transform.*

**Methods**

The MoS$_2$ was grown on an oxidized silicon wafer (300 nm SiO$_2$ on Si) by TAC of pre-deposited Mo. The resulting MoS$_2$ had a thickness of approximately 4 nm, which should theoretically result in an electronic band gap of E$_g$ = 1.45 eV [27]. Details about the process tool and synthesis parameters are given in the Supporting Information. The layers were subsequently transferred by a conventional wet method using spin-coated polymethyl methacrylate (PMMA) as a support for the MoS$_2$. The 2D film was separated from the growth substrate by etching the Si/SiO$_2$ in potassium hydroxide. The supported MoS$_2$ was transferred onto deionized water for cleaning and fished out with the target substrate, n-doped a-Si:H (n-a-Si:H) on a gold-plated Si/SiO$_2$ wafer (cf. Figure1b). The PMMA was thoroughly removed by a multistep process using ultraclean solvents. Next, an intrinsic a-Si:H (i-a-Si:H) absorber and a p-doped a-Si:H (p-a-Si:H) were grown onto the MoS$_2$ in a conventional plasma-enhanced CVD (PE-CVD) process in a multi-chamber vacuum system. The Supporting Information and Ref. [28] provide detailed process parameters. The MoS$_2$ films intentionally did not cover the entire sample area, so that MoS$_2$-free regular photodiodes were fabricated next to the IR-sensitive MoS$_2$/a-Si:H detectors. Aluminum-doped zinc-oxide (AZO) was deposited *in-situ* to avoid unwanted contamination at the semiconductor-/contact interface. The single devices were finally separated by selective ion etching of the front contact and the a-Si:H pin-diode, including the embedded FL-MoS$_2$.

After the MoS$_2$ transfer onto the substrate and prior to further PE-CVD deposition of a-Si:H, Raman analyses were conducted to examine the MoS$_2$ quality and to confirm its thickness. The spectra were recorded using a confocal WITec Alpha 300R Raman microscope equipped with a 532 nm laser. The signal was collected through a 100× objective and dispersed using an 1800 g/mm grating. A Peltier cooled charge-coupled device (Andor DU401 BV) was used for the spectrum detection. During the Raman measurements, the laser power was kept constant at 0.5 mW.

The process integration, i.e. the quality and homogeneity of the thin film stacks, the roughness of the embedded MoS$_2$ and the wetting and growth of the a-Si:H on the FL-MoS$_2$, was investigated by scanning and transmission electron microscopy (SEM, TEM) in conjunction with focused ion-beam (FIB) cross-sectioning. An FEI Strata dual-beam FIB/SEM was utilized for SEM and site-specific TEM sample preparation by regular FIB lift-out. High-resolution TEM (HRTEM), scanning TEM (STEM) and energy-dispersive X-ray spectroscopy (EDS) at 200 kV were carried out using a FEI Talos F200A with a large-area ChemiSTEM$^{TM}$ EDS detector.

Electrical characterization comprised current-voltage measurements (I-V) with and without standardized AM1.5 illumination and was performed using a Suss microprobe station connected to a Keithley 4200-SCS parameter analyzer. For spectrally resolved photoresponse measurements, a monochromator (Acton Research, tungsten-halogen lamp) was utilized to sample narrow wavelength bands of 10 nm in the range of 400 nm to 2120 nm.

The photocurrent signals were converted using a FEMTO DLCPA-200 I-V converter and amplifier and acquired via lock-in technique (Princeton 5210). The optical setup was calibrated using a reference c-Si photodetector (S1337-33BQ, Hamamatsu) for wavelengths below 1100 nm and an InGaAs photodetector (12182-030k, Hamamatsu) for wavelengths up to 2120 nm (cutoff wavelength) to extract absolute spectral responsivity values. All electrical and optical measurements were performed under ambient conditions.

## Results and discussion

Figure1a and c demonstrate the successful integration of large-area FL-MoS$_2$ into periodic arrays of a-Si:H pin-diode photodetectors. The targeted MoS$_2$ thickness of ~4 nm as well as the interlayer spacing were verified by HR(S)TEM (cf. Figure1d) and coincide with the expected thickness increase of the initial metallic Mo by a factor of three [29]. After the repeated fabrication of entire MoS$_2$/a-Si:H photodiodes (cf. Supporting Information, *Methods*), the intended layer sequence of the devices was verified by site-specific cross-sectional S/TEM analyses (cf. Figure1c, d). The S/TEM images confirm excellent homogeneity of the FL-MoS$_2$ (cf. Figure1c, d) and exceptional wetting of the i-a-Si:H on the MoS$_2$, a prerequisite for the successful integration. The devices show minimal interface roughness, in particular in close vicinity to the embedded FL-MoS$_2$. Consequently, the stacks exhibit minimal thickness variations (detailed information is provided in the Supporting Information). For vertical devices, significant PMMA contamination after the transfer of the thin-film semiconductor is expected to negatively affect the i-a-Si:H nucleation and growth. More importantly, it would govern the charge transfer to, as well as the transport through, the pin-junction. However, nanoscopic TEM characterization did not reveal such residuals indicating the thorough removal of the PMMA. This demonstrates that routine integration of a 2D layered semiconductor like FL-MoS$_2$ in regular semiconductor-processing schemes is feasible and can considerably extend the functionality and performance of current devices. In this case, functional arrays of FL-MoS$_2$ based devices up to 0.25 mm² on a total area of 4 cm² were realized.

J-V curves of an MoS$_2$-based photodiode and a reference a-Si:H pin-diode are presented in Figure 2a. For both detector types, we extracted ideality factors of $n \approx 3$ from the dark-current density ($j_{dark}$) measurements. An ideality factor of n = 2 for the a-Si:H pin-diode was extracted under AM1.5 illumination, which compares well to the predicted values in the range of 1.4-2 [30]. Under illumination, the shape of the J-V characteristic of the MoS$_2$/a-Si:H heterostructure changes substantially at low bias voltages so that a substantiated extraction of the ideality factor is almost impossible. The photocurrent decrease might be due to oxide formation at the MoS$_2$ interfaces most likely originating from the wet transfer. The presence of oxygen has been confirmed by EDX line-scan measurements (cf. *Figure 1c and S3*). Local oxidation of the a-Si:H presumably reduces the vertical charge transfer along the heterojunction and leads to an additional, parasitic series resistance. However, a decade change in current by increasing the external bias from 0.6 V to 1.5 V would result in n ≈ 15.. The bare a-Si:H detector exhibits an exceptional high dynamic range DR $= 20 \cdot \log\left(\frac{j_{photo}}{j_{dark}}\right)$ of ≥100 dB at low bias voltages. In the low reverse bias regime, the dynamic of the MoS$_2$/a-Si:H diodes likewise reach ≥ 80 dB which is comparable to the performance of common a-Si:H based mono- or multicolor photodetectors [31,32]. The slight reduction in dynamic and drop in the ideality factor may be explained by additional series resistances due to charge transfer perpendicular to the MoS$_2$ layers and/or by charge recombination at additional defects and interfaces. Under larger reverse bias conditions of V$_{bias}$ < -2 V, both detectors perform nearly identical and exhibit good dynamic ranges of around 60 dB at -6 V. Even after several months, the DR remains almost constant (cf. *Figure S3, S4*) demonstrating that capping the MoS$_2$ by a-Si:H conserves its properties.

*Figure 2: a) J-V characteristics of MoS$_2$/a-Si:H and a-Si:H reference pin-diodes under AM 1.5, b) R$_\lambda$ of both detector types under reverse bias (-6 V, measured up to 2120 nm ≡ 0.58 eV), left inset: transition region at 800 nm, right inset: responsivity maxima in the visible range for both detector types, c) bias-dependent R$_\lambda$ of a representative MoS$_2$/a-Si:H photodetector under reverse bias of -4 V and -5 V, d) demonstration of clear bias-dependent responsivity switching above 800 nm.*

The absolute photoresponsivities R$_\lambda$ of a MoS$_2$/a-Si:H heterostructure and a typical a-Si:H reference pin-photodiode are shown in Figure 2b for an applied bias of -6 V. Although absolute photoresponsivities of the reference devices slightly vary due to slight inhomogeneities of the AZO contact deposition, the R$_\lambda$ of all a-Si:H pin-detectors is qualitatively identical (cf. Figure 2b,c), confirming reproducible a-Si:H fabrication conditions. They reach an almost bias-independent global maximum of around 320 mAW$^{-1}$ at 520 nm (V$_{bias}$ = -6 V) (cf. Figure 2b) corresponding to an external quantum efficiency (EQE) of about 75 %. This performance compares well to other a-Si:H pin-detectors [33,34]. Being insensitive for infrared light due to the a-Si:H bandgap of 1.65 eV (cf. *Figure S8b*), R$_\lambda$ of these reference detectors vanishes above 800 nm. By optical simulations it has been verified that the local maxima at 600 and 650 nm result from interference effects between the different layers (cf. *Figure S9*). These maxima shift in the MoS$_2$/a-Si:H detectors (Figure 2b, right inset) as the interference conditions change due to increased device thickness (cf. *Figure S9b,c*). Despite the very small mass fraction compared to the i-a-Si:H absorber, the embedded MoS$_2$ effectively extends the spectral detection bandwidth of our MoS$_2$/a-Si:H pin-diodes in a broad range far into the mid-IR (cf. Figure 2b). Although the band gap of the grown MoS$_2$ has been determined by spectroscopic ellipsometry to be E$_g$ = 1.45 eV (cf. *Figure S8a*), systematic absorption far below that inherent optical bandgap occurs. The presence of the MoS$_2$ is thus not sufficient to explain this repeatable observation. Instead, the absorption may be due to a high density of defect states in the sulfurized material and/or at the interfaces to the a-Si:H. This hypothesis is in line with experiments on pure MoS$_2$, where it was concluded that sulfur vacancies in the MoS$_2$ lattice reduce the bandgap and lead to a metallic behavior of MoS$_2$ above a certain level of disorder [15]. The IR photoresponsivities of different MoS$_2$/a-Si:H photodetectors above 750 nm were found to follow a general trend (cf. *Figure S7*), although the detailed wavelength dependencies vary. This may be attributed to local variability of the MoS$_2$ (and thus the local interface to the a-Si:H), similar to the shifts in interference maxima in the visible range.

A key parameter to evaluate the photon-to-charge conversion for specific wavelengths is the detectivity $D^* = \frac{R_\lambda}{\sqrt{2 \cdot q \cdot j_{dark}}}$ [35]. Here, V$_{bias}$ allows a facile tuning of the dark-current density $j_{dark}$, R$_\lambda$ and therefore D*. The MoS$_2$/a-Si:H heterojunction diodes with FL-MoS$_2$ reach a mean R$_\lambda$ of around 50 mAW$^{-1}$, a respective D* of 2·10$^{10}$ Jones, and an EQE of nearly 3 % in the IR at 2120 nm ≡ 0.58 eV. The mean EQE in the IR is larger than 1 % above 1100 nm at -6 V bias voltage (cf. Figure 2b).

An interesting feature of our MoS$_2$/a-Si:H detectors is their capability to be easily switched between a limited optical response in the visible regime and an extended response comprising the visible and IR (cf. Figure 2c, d). By applying a specific external voltage to the MoS$_2$/a-Si:H heterojunction (cf. Figure 2d), charge from the MoS$_2$ contributes to the total spectral responsivity. We assume, that - as a result of the MoS$_2$ integration - defect states within the MoS$_2$ itself or interface states might cause defect-to-band transitions at wavelength below the initial MoS$_2$ bandgap. It is already known that sulfur vacancies introduce additional defect levels within the MoS$_2$ bandgap [36,37]. It is possible, mobile OH- ions in MoS$_2$ interact with interface states [38] causing charge screening effects above a certain external bias. We believe that other TMDs, e.g. PtSe$_2$ [39], synthesized by the TAC approach might have similarly high defect densities and might also produce similar but not identical effects to those presented in our work. Due to the local variation of the MoS$_2$, the bias voltage required to activate IR sensitivity varies between samples. However, our detectors facilitate different operation modes. They can be used for panchromatic measurements at higher reverse bias to detect the sum of incoming visible plus IR contributions. At low reverse bias, the clear bias dependency facilitates the distinction between visible and IR components (e.g., $I_{IR} = I_{-5V} - I_{-4V}$, cf. Figure 2c) using just a single pixel detector.

Table 1 compares our results with those of previous studies on MoS$_2$ based photodetectors with an extended responsivity in the VIS and the IR range. The listed devices were realized in more simple lateral geometric configurations and are based on exfoliated flakes, i.e. limited in size and scalability. In addition, the IR response in most of these studies was achieved through specific band gap engineering in 2D heterostructures, plasmonics or nanoparticles. Esmaeili-Rad et al. [40] deposited a-Si:H on a mechanically exfoliated FL-MoS$_2$ flake, but no IR photoresponse was measured for the lateral MoS$_2$/a-Si:H heterojunction photodetector. A scalable IR sensitive lateral quasi-bulk MoS$_2$ photoconductor has been demonstrated by Xie et al. [15] with slightly lower R$_\lambda$ and D* under high power illumination. However, these devices were not tunable (cf. Table 1). Our approach is therefore unique that it utilizes standard semiconductor processes and a scalable MoS$_2$ synthesis process and results in tunable IR sensitive photodetectors. The vertical detector arrays are based on the mature a-Si:H platform and show long-term functionality (cf. , *Figures S5-7*).

Table 1: Comparison of MoS$_2$-based photonic devices with spectral responsivity beyond the MoS$_2$ bandgap.

| Ref. | Detector type<br>Assembly<br>Operation conditions | Device dimensions<br>Active area | Radiant power#<br>/ W | $\lambda_{reference}$<br>/ nm | $R_\lambda$<br>/ AW$^{-1}$ | D*<br>/ Jones | Spectral range<br>/ nm | Thickness of 2D-material |
|---|---|---|---|---|---|---|---|---|
| This work | **Vertical photodiode**<br>MoS$_2$/a-Si:H pin-junction<br>$V_{bias}$ = -6 V | **Any shape**<br>**100-500 µm**<br>2 x 2 cm² | 0.7<br>x 10$^{-6}$ | 2120 | 50<br>x 10$^{-3}$ | 2<br>x 10$^{10}$ | 400–2120<br>(minimum) | 5–6 layers<br>(~4 nm) |
| Deng et al. 2018 [41] | **Lateral phototransistor**<br>MoS$_2$/graphene (Gr)/MoS$_2$<br>$V_G$ = 0 V, $V_{DS}$ = 0.5 V | ~ 100 µm² | 3<br>x 10$^{-7}$ | 1400 | ~6<br>x 10$^{-6}$ | ~2<br>x 10$^{7}$ | 532-1600 | 1 layer Gr<br>MoS$_2$ (N.A.) |
| Xie et al. 2017 [15] | **Lateral photoconductor**<br>MoS$_2$<br>V = 10 V | **60 µm x<br>2 mm**<br>Ø = 25 mm | 8<br>x 10$^{-3}$ | 2111 | 32<br>x 10$^{-3}$ | <10$^{9}$ | 400–2717 | 73 layers |
| Huo et al. 2017 [42] | **Lateral photoconductor**<br>MoS$_2$ + HgTe quantum dots<br>$V_G$ = 15V, $V_{DS}$ = 1V | N.A.<br>(< 10 µm²) | 0.7<br>x 10$^{-12}$ | 2000 | 10$^{5}$ | 10$^{12}$ | 600–2100 | few layers |
| Ye et al. 2016 [43] | **Lateral photodetector**<br>BP/MoS$_2$ pn-junction<br>$V_G$ = 60 V, $V_{DS}$ = 3 V | N.A.<br>(~30 µm²) | 1.5<br>x 10$^{-6}$ | 1550 | 30<br>x 10$^{-3}$ | 2<br>x 10$^{9}$ | 1550 | 22 nm BP<br>12 nm MoS$_2$ |
| Zhang et al. 2016 [44] | **Lateral photoconductor**<br>MoTe$_2$/MoS$_2$ pn-junction<br>($V_{DS}$ ~ 0.8 V) | N.A.<br>(~50 µm²) | 0.5<br>x 10$^{-3}$ | 1550 | 17<br>x 10$^{-6}$ | --- | 1550 | 1 layer MoTe$_2$<br>1 layer MoS$_2$ |
| Kufer et al. 2015 [12] | **Lateral photoconductor**<br>MoS$_2$ + PbS quantum dots<br>$V_G$ = 100 V, $V_{DS}$ = 1 V | 12/15 µm² | 2<br>x 10$^{-12}$ | 1000 | 2.2 | 7<br>x 10$^{14}$ | 550–1170 | >2 layers |
| Wang et al. 2015 [45] | **Lateral photoconductor**<br>MoS$_2$ + plasmonic antenna<br>$V_{bias}$ = 0.8 V | N.A. | 1.5<br>x 10$^{-7}$ | 1070 | 5.2 | --- | 1000–1250 | 2 layers |
| Yu et al. 2014 [46] | **Lateral photoconductor**<br>Dye-sensitized MoS$_2$<br>$V_G$ = 0 V, $V_{DS}$ 5 = V | ~80 µm² | 1<br>x 10$^{-3}$ | 1000 | <1<br>x 10$^{-6}$ | 10$^{2}$ | 400–1000 | 1 layer |

#All former studies utilized laser illumination; here standardized/natural illumination is applied; BP: black phosphorous

**Conclusion**

This work demonstrates the systematic, reproducible, and scalable integration of large-area TAC $MoS_2$ as a 2D layered semiconductor into the mature a-Si:H thin-film technology for vertical photodiodes. We show the large-scale synthesis and transfer of homogeneous $MoS_2$ as well as the subsequent growth of a-Si:H photodiodes by PE-CVD. The $MoS_2$ extends the detector bandwidth from the optical range far into the mid-IR (minimum up to 2120 nm) with an almost constant responsivity above 800 nm. The new $MoS_2$/a-Si:H photodiodes exhibit good performance (DR = ~60 dB at -6 V, $R_\lambda$ ~50 mAW$^{-1}$/ D* = ~2·10$^{10}$ Jones / EQE ~3 % at 2120 nm under ambient conditions) in conjunction with an excellent long-term stability. The detectors realized in this work combine several advantages:

I)  they are suited for various applications where a tunable photodetection in the visible regime and beyond is needed,

II) they are easily operated under ambient conditions at room temperature combined with long-term stability, and, most importantly,

III) they are based on a scalable, mature optoelectronic thin-film platform allowing for large-scale integration in detector arrays and making the integration of additional fabrication steps reliable, easy and economic.

Prospectively, such detector arrays in combination with state-of-the-art read-out electronics, pave the way for the realization of pixelated, wavelength-selective sensors with a photoresponse in a wide spectral range. Further responsivity enhancement by bandgap engineering like the defined introduction of defects, the additional implementation of other 2D materials and/or the facile tuning of the a-Si:H composition and thickness might even enable novel a-Si:H multi-spectral photodetectors or high-efficient solar cells covering an even broader spectrum from the ultraviolet (UV)/VIS to the IR.

**Acknowledgement**

The authors gratefully acknowledge the preparation of the FIB-lamellae by Dr. Erich Müller from the Laboratory for Electron Microscopy (Prof. Dr. Dagmar Gerthsen)/Karlsruhe Institute of Technology KIT. Dr. Matthias Adlung from the Institute for Inorganic Chemistry (Prof. Dr. Claudia Wickleder)/University of Siegen is given credit for absorption spectra measurements. Part of this work was performed at the Micro- and Nanoanalytics Facility (MNaF) of the University of Siegen. G. S. D. acknowledges the support of SFI under Contract No. 12/RC/2278 and PI_15/IA/3131 as well as Graphene Flagship under Contract 785219. N. M. acknowledges support from SFI through 15/SIRG/3329.

**Supporting Information**

Proposed band structure, fabrication parameters, EDX line-scan, ideality factor extraction, $MoS_2$ Raman spectra, long-term J-V and $R_\lambda$ stability, a-Si:H and $MoS_2$ absorption coefficients, optical simulation

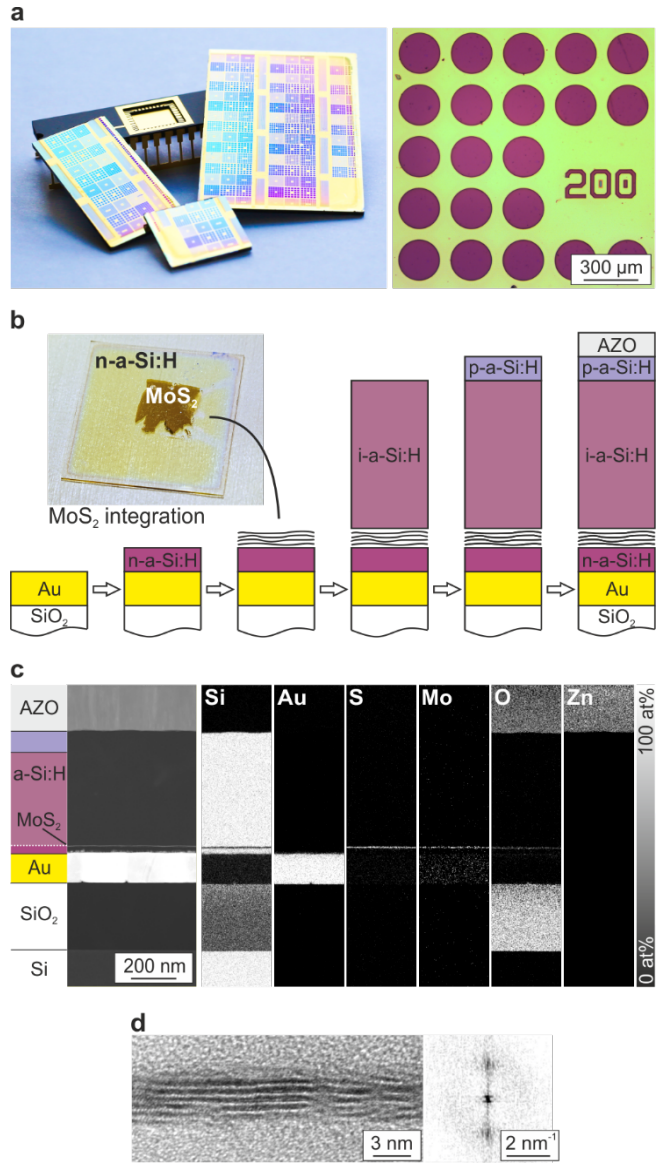

*Figure 3: a) left: photograph of chips with MoS₂/a-Si:H and reference photodiodes, right: microphotograph of circular 200 μm MoS₂/a-Si:H detectors, b) fabrication scheme including MoS₂ integration, c) cross-sectional STEM micrograph with corresponding EDS element maps, d) HRTEM detail of embedded MoS₂ thin film with corresponding fast Fourier transform.*

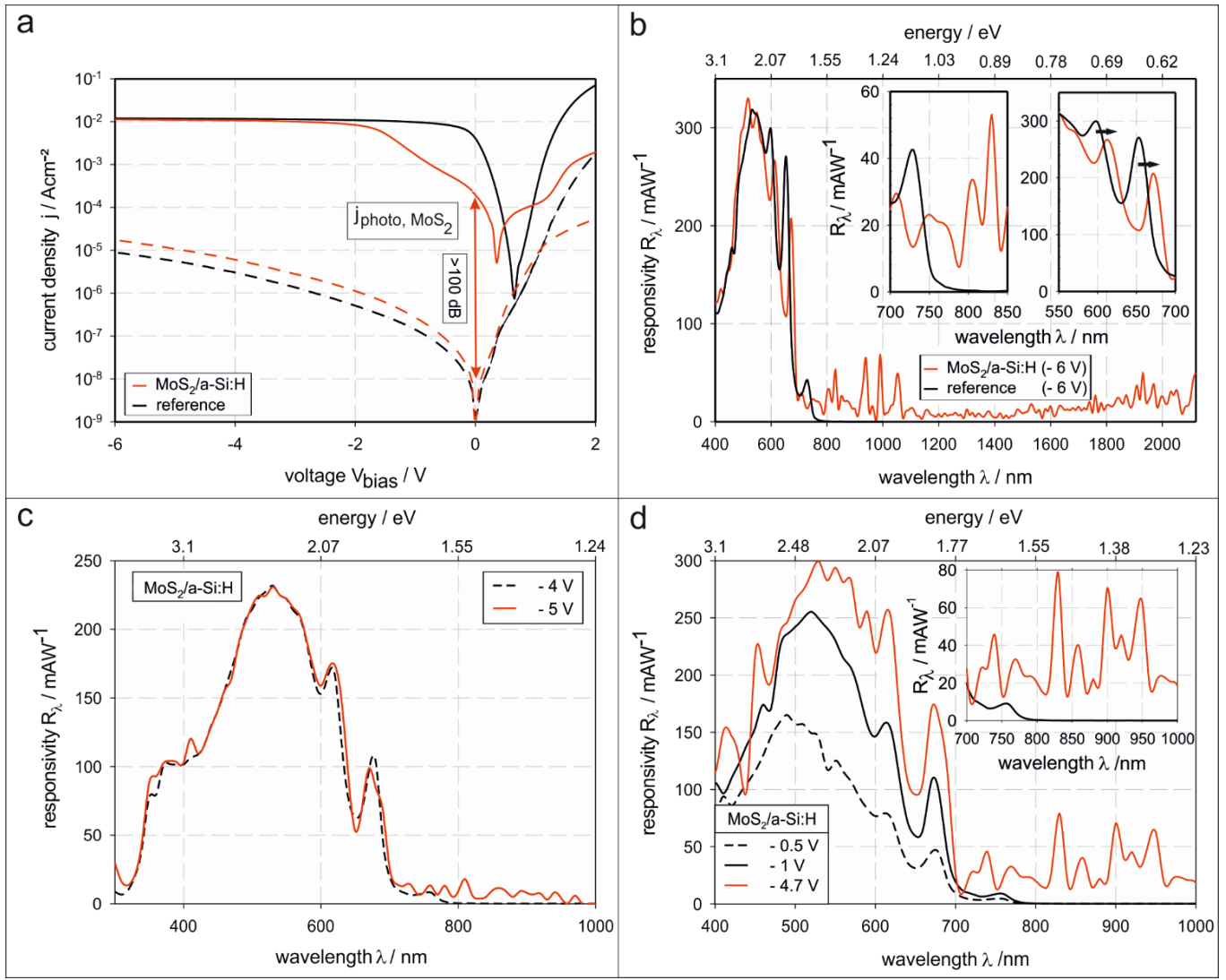

*Figure 4: a) J-V characteristics of MoS₂/a-Si:H and a-Si:H reference pin-diodes under AM 1.5, b) $R_\lambda$ of both detector types under reverse bias (-6 V, measured up to 2120 nm ≡ 0.58 eV), left inset: transition region at 800 nm, right inset: responsivity maxima in the visible range for both detector types, c) bias-dependent $R_\lambda$ of a representative MoS₂/a-Si:H photodetector under reverse bias of -4 V and -5 V, d) demonstration of clear bias-dependent responsivity switching above 800 nm.*